\documentstyle[12pt,psfig]{article}	 
\begin{document}

\title{ Electron--vibration coupling constants in positively charged 
fullerene.}

\author{Nicola Manini$^{1,2}$\thanks{E-mail: nicola.manini@mi.infm.it},
        Andrea Dal Corso$^{1,2}$\thanks{E-mail: dalcorso@sissa.it},
\\
        Michele Fabrizio$^{1,2,3}$\thanks{E-mail: fabrizio@sissa.it},
    and Erio Tosatti$^{1,2,3}$\thanks{E-mail: tosatti@sissa.it}
\\ \it 
$^1$ Istituto Nazionale di Fisica della Materia (INFM), 
\\ \it
Unit\`a Trieste SISSA,
\\ \it 
$^2$ International School for Advanced Studies (SISSA),
\\ \it
Via Beirut 4, I-34013 Trieste, Italy
\\ \it 
$^3$ International Centre for Theoretical Physics (ICTP), 
\\ \it
P.O. Box 586, I-34014 Trieste, Italy
}

\date{April 10, 2001}
\maketitle

\begin{abstract}
Recent experiments have shown that C$_{60}$ can be positively
field-doped. In that state, fullerene exhibits a higher resistivity and a
higher superconducting temperature than the corresponding negatively doped
state. A strong intramolecular hole-phonon coupling, connected with the
Jahn-Teller effect of the isolated positive ion, is expected to be
important for both properties, but the actual coupling strengths are so far
unknown.
Based on density functional calculations, we determine the linear couplings
of the two $a_g$, six $g_g$, and eight $h_g$ vibrational modes to the $H_u$
HOMO level of the C$_{60}$ molecule.
The couplings predict a $D_5$ distortion, and an $H_u$ vibronic ground
state for C$_{60}^+$.
They are also used to generate the dimensionless coupling constant
$\lambda$ which controls the superconductivity and the phonon contribution
to the electrical resistivity in the crystalline phase. We find that
$\lambda$ is 1.4 times larger in positively-charged C$_{60}$ than in the
negatively-doped case.
These results are discussed in the context of the available transport data
and superconducting temperatures. The role of higher orbital degeneracy 
in superconductivity is also addressed.
 
\end{abstract}


\section{Introduction}

Recent breakthrough experiments (Sch\"on {\it et al.} 2000) have shown that
a monolayer of C$_{60}$ can be positively field-doped, a goal hardly
realized chemically so far (Datars and Ummat 1995). In that state,
fullerene exhibits a higher resistivity than for negative doping, and
becomes superconducting with critical temperatures that can exceed 50~K,
about a factor 5 higher than the corresponding negative field-doped state.
The general belief is that superconductivity in the fullerenes is related
to a strong intra-molecular electron-phonon coupling, connected with the
Jahn-Teller effect of the isolated ion (Antropov {\it et al.} 1993). Unlike
the negative doping case, where both calculations
(Antropov {\it et al.} 1993, Varma {\it et al.} 1991, Lannoo {\it et al.}
1991) and fits to data (Gunnarsson {\it et al.} 1995) exist, no
quantitative evaluation of the actual Jahn-Teller coupling strengths is so
far available for the positive fullerene ions.

In this work we undertake the task of determining the electron-vibration
linear couplings for the $H_u$ HOMO level of the C$_{60}$ molecule, along
with that of spelling out some of the consequences for resistivity and for
superconductivity.   For that purpose we use density functional electronic
structure calculations, yielding accurate molecular vibration frequencies
and eigenvectors for the two $a_g$, six $g_g$, and eight $h_g$ modes that
couple linearly to the electronic $H_u$ state.  Knowing the form of the
Jahn-Teller coupling matrices, we distort the molecule and extract the
coupling constants from the calculated level shifts and splittings. 
As a parallel check, we repeat a similar calculation for the negative,
electron doping case, where the $a_g$ and $h_g$ modes couple
linearly to the $T_{1u}$ LUMO level of the C$_{60}$ molecule.
The couplings obtained for negatively doped C$_{60}$ are rather similar to
those that can be found in the literature, and just represent a fresher,
state-of-the-art theoretical determination.
The hole-vibration couplings of positively doped C$_{60}$ are new, and can
be put to use in a variety of manners, including predicting or explaining
properties of molecular ions, such as photoemission (Br\"uhwiler {\it et
al.} 1997) and IR/Raman spectra.
That is a task that we propose to consider in the near future.

The couplings obtained can also be used to determine the 
dimensionless electron-phonon coupling constant $\lambda$ relevant for the 
superconductivity as well as for the vibron contribution to the 
high temperature $T$-linear resistivity in the crystalline phase. 
Comparing values for positive and negative doping we find that for positive
doping $\lambda$ is a factor 1.4 larger than for negative doping.  
These results provide a starting point for a discussion and comparison 
with the experimental findings.

This paper is organized as follows: the notation is set up in
Sect.~\ref{ham:sec}; the ab-initio calculation and results for the
molecular ion are described is Sect.~\ref{dft:sec}; in
Sect.~\ref{solid:sec} we sketch the calculation of the resistivity in a
band-degenerate case; Sect.~\ref{superc:sec} presents a formulation 
for superconductivity in that case; Sect.~\ref{discussion:sec} contains
comparisons and discussion of experimental data.

\section{The Jahn-Teller Hamiltonian}
\label{ham:sec}

Several theoretical papers (Ceulemans and Fowler 1990, De Los Rios {\it et
al.} 1996, Moate{\it et al.} 1996, Moate {\it et al.} 1997, Manini and De
Los Rios 2000) formulate the $H \otimes (a+g+h)$ Jahn-Teller (JT) and
dynamical JT problem -- describing a hole in molecular C$_{60}$ -- though
with different notations and conventions.
For ease of comparison, it is therefore useful to set up explicitly the
conventions we use in the present calculation.

The analogies of the icosahedral $H \otimes (a+g+h)$ JT coupling to a
spherical ${\cal D}^{(2)}\otimes \left({\cal D}^{(0+)} + {\cal D}^{(2+)} +
{\cal D}^{(4+)}\right)$ model were exploited in earlier work, where among
other things it was shown that surprisingly, and depending on numbers, this
dynamical JT problem may or may not possess a Berry phase (De Los
Rios {\it et al.} 1996, De Los Rios and Manini 1997, Manini and De Los
Rios 1998).
The representations ${\cal D}^{(L\pm)}$ of $O(3)$ map into
representations of the icosahedral group ${\cal I}_h$ as follows
(Altmann  and Herzig 1994):
${\cal D}^{(0\pm)} \longrightarrow a_{g/u}$,
${\cal D}^{(1\pm)} \longrightarrow t_{1\,g/u}$,
${\cal D}^{(2\pm)} \longrightarrow h_{g/u}$,
${\cal D}^{(3\pm)} \longrightarrow t_{2\,g/u}\oplus h_{g/u}$,
${\cal D}^{(4\pm)} \longrightarrow g_{g/u}\oplus h_{g/u}$,
and so on.
This means that the decomposition of the symmetric part of the tensor
product
\begin{equation}
\{ {\cal D}^{(2-)} \otimes {\cal D}^{(2-)}{\}}_s = 
                {\cal D}^{(0+)} \oplus {\cal D}^{(2+)} \oplus {\cal
                D}^{(4+)}
\label{so3}
\end{equation}
becomes in icosahedral language
\begin{equation}
\{ h_u \otimes h_u {\}}_s = 
                a_g \oplus h_g \oplus (g_g \oplus h_g) \ .
\label{ihdecomp}
\end{equation}
$a_g$ and $g_g$ appear only once, while the $h_g$ representation
modes appears twice
in this decomposition: one represents a genuine quadrupolar ${\cal
D}^{(2+)}$ state, while the other one derives from a ${\cal D}^{(4+)}$
representation.
Though this $O(3)$ picture is suggestive (Ceulemans {\it et al.} 1994),
clearly a quantitative description of C$_{60}^{n+}$ ions had better involve
icosahedral symmetry from the beginning.

In the icosahedral group indeed in the $h_u \otimes h_u$ tensor product,
the $h_g$ representation appears twice.
This reflects the {\em non-simple reducibility} of the icosahedral symmetry
group.
Accordingly, Butler 1981 provides two independent
sets of Clebsch-Gordan (CG) coefficients
\begin{equation}
^hC_{\mu,\nu}^{m~[r]}\; \equiv\; \langle H,\mu;H,\nu|h,m\rangle^{[r]}
\end{equation}
which couple an $H$ electronic state (quadratically) with an $h$
vibrational mode (linearly) to give a scalar.
Each set of coefficients is identified by a multiplicity index $r=1,2$.
Since the two $h$ states labeled $r=1,2$ are symmetry-wise
indistinguishable, the choice of these two sets of coefficients is
perfectly arbitrary, as long as they are kept orthogonal to each other.
This arbitrariness is the source of the different notations taken in the
literature of this field.
Here, we stick to Butler's choice (Butler 1981), which is basically
equivalent to Ceulemans' convention (Fowler and Ceulemans 1985).
Also, we label the states within a degenerate multiplet by the labels of
the subgroup chain ${\cal I}_h \supset D_5 \supset C_5$.
For brevity, we indicate only the $C_5$ index $m$ ($m=0$ for $a_g$,
$m=\pm 1,\pm 2$ for $g_g$ and $m=-2,\dots,2$ for $h_g$ states) in the
labeling of states since, for the representations relevant to our problem,
the $D_5$ label is just the absolute value of $m$.

Given the tabulated CG coefficients, it is necessary for generality to
consider a linear combination
\begin{equation}
	^hC_{\mu,\nu}^m\left(\alpha\right)\equiv 
	\cos\alpha ~^hC_{\mu,\nu}^{m~[1]} + \sin\alpha ~^hC_{\mu,\nu}^{m~[2]} 
\end{equation}
of the two sets.
The coefficient $^hC_{\mu,\nu}^m\left(\alpha\right)$ coincides with Butler's
$r=1$ and $r=2$ values for $\alpha=0$ and $\alpha=\frac \pi 2$ respectively.
Different  values of $\alpha$ can be compared with the conventions of previous
authors.
For example, 
$\alpha=-\arctan\left( 3 / \sqrt 5 \right) \approx -53.3^\circ$ 
is the case studied by De Los Rios {\it et al.} 1996 (where the ${\cal
I}_h$ CG coefficient becomes equivalent to the spherical $\langle
2,\mu;2,\nu|2,m\rangle$); $\alpha=\pi/2$ by Moate {\it et al.} 1996; and
$\alpha=0,\pi/2$ by Ceulemans and Fowler 1990, where these cases are
indicated as $h_b$ and $h_a$ respectively.
The $\alpha$-dependence of these CG coefficients indicates that -- unlike,
for example, cubic symmetry -- belonging to the
$h_g$ group representation in icosahedral symmetry does not determine 
completely the form of the JT
coupling. The mixing angle $\alpha$ is also needed for that.
In the present case of fullerene, each $h_g$ distortion mode is thus
characterized not only by its frequency and scalar coupling, but also
by its specific mixing angle $-\pi/2\leq\alpha\leq\pi/2$.  

The basic linear Jahn-Teller Hamiltonian for the $H_u\otimes (a_g+g_g+h_g)$
model is conveniently divided into:
\begin{equation}
H = \sum_{\tau=a_g,g_g,h_g} \sum_i^{n_{\rm modes}(\tau)} 
\left[ H_{\rm harm}^\tau(\hbar \omega_{\tau i},\vec P_{\tau i},\vec Q_{\tau i})
+ H_{\rm e-v}^\tau 
(g_{\tau i} \hbar \omega_{\tau i},\alpha_{\tau i},\vec Q_{\tau i}) 
\right]\;.
\label{hamiltonian1mode}
\end{equation}
The first term describes the linearly-coupled vibrations 
in the  harmonic approximation, 
\begin{equation}
H_{\rm harm}^\tau(\hbar \omega,\vec P,\vec Q)= 
\frac{ \hbar \omega }{2} \sum_{m} (P_m^2 +Q_m^2) \,,
\end{equation}
while the second term is the linear coupling to each mode:
\begin{equation}
H_{\rm e-v}^\tau(g \hbar \omega ,\alpha,\vec Q)=
\frac {g \hbar \omega}2
\sum_{m\,\mu\,\nu} Q_{m}
\;
c^\dagger_{\mu} c_{-\nu} 
\;
^{\tau}C_{\mu,\nu}^m\left(\alpha\right) \; .
\label{interaction hamiltonian}
\end{equation}
Here, of course, the $\alpha$ dependence is relevant only for the
$\tau=h_g$ case.
The distortion coordinates $Q_{\tau i m}$ (with conjugate momentum $P_{\tau
i m}$) are dimensionless, being measured in units of $x_0(\omega_{\tau i})
=(\hbar / m_{\rm C}\; \omega_{\tau i})^{1/2}$.
The operator $c^\dagger_{\mu}$ creates an electron in 
orbital $\mu(=-2,\dots,2)$ of the HOMO $H_u$ shell.

Naturally this form of the coupling Hamiltonian is such that 
each term represents pertinent irreducible representation combinations 
that are totally symmetrical, i.e., scalars, under the icosahedral group.
For future applications, it will provide a convenient form both for
perturbative calculations (small $g$ values) and as a starting point for
numerical diagonalization methods, such as the Lanczos technique.
Here, however, we restrict ourselves to a study of the classical molecular
distortions. For that, it is more convenient (Manini and De Los Rios 2000)
to switch to a {\em real representation} of the vibrational degrees of
freedom and orbitals.

To that end, we apply two standard unitary transformations (Manini and De
Los Rios 2000), one to the electronic and the other to the vibrational
degrees of freedom.
We define a new set of electronic operators, $d_m$ (and consequently their
Hermitian conjugates $d_m^\dagger$), as
\begin{eqnarray}
\label{ele trans}
c_0&=&d_0 \\
\pmatrix{ c_m \cr c_{-m} }
&=& \frac 1{\sqrt{2}}
\left(\begin{array}{cc} 1 & i \\ 1 & -i \end{array} \right)
\pmatrix{ d_m \cr d_{-m} }, \  m=1,2~. \nonumber
\end{eqnarray}
This transformation leaves unchanged the (diagonal) coupling to the $a_g$
modes, which takes the final form:
\begin{equation}
H_{\rm e-v}^{a_g}(g \hbar \omega,q)=
\frac {g \hbar \omega}2 \; q \;
\sum_{\mu\,\nu} d^\dagger_{\mu} d_{\nu} V^{a_g}_{\mu\nu} \ ,
\label{ag interaction}
\end{equation}
with
$
V^{a_g}_{\mu\nu}=\delta_{\mu\nu}
$
and $q=Q_0$.

The second (similar) transformation is applied to the vibrational
coordinates of the $g_g$ and $h_g$ modes:
\begin{eqnarray}
\label{coordinate}
Q_0 & = & q_0  \ \ \  \ \ \  \ \ \  \ \ \quad (h_g\ {\rm modes\ only})
\\
\pmatrix{Q_m \cr Q_{-m} }
&=&
\frac {(-1)^m}{\sqrt{2}}
\left(\begin{array}{cc} 1 & i  \\ 1 & -i \end{array} \right)
\pmatrix{ q_m \cr q_{-m} } \  (m=1,2)\ . \nonumber
\end{eqnarray}
The harmonic part is left unchanged by this transformation, while 
the interaction is transformed into 
\begin{equation}
H_{\rm e-v}^\tau(g \hbar \omega ,\alpha,\vec q )=
\frac {g \hbar \omega}2
\sum_m q_m \sum_{\mu\,\nu}  d^\dagger_{\mu} d_{\nu}
V^{\tau\;(m)}_{\mu\,\nu}(\alpha) \ \ \ \ (\tau=g_g,h_g)\; .
\label{real interaction Hamiltonian}
\end{equation}
The $5\times 5$ coupling matrices $ {\bf V}^{\tau\;(m)}$ are combinations
of the CG coefficients.  Their explicit expressions [we use the shorthand
$s$ for $\sqrt{3}$ and omit the explicit indication of dependence 
${\bf V}^{h_g\;(m)}(\alpha)$] are the following:
$${\bf V}^{g_g\;(-2)}=s^{-1}\;
 \left(
\matrix{ 0 & 0 & 0 & - \frac{1}{4} \
  & 0 \cr 0 & 0 & 0 & 1 & - \frac{1}{4} \
  \cr 0 & 0 & 0 & 0 & \frac{s}{2} \cr - \frac{1}
     {4}   & 1 & 0 & 0 & 0 \cr 0 & - \frac{1}{4} \
  & \frac{s}{2} & 0 & 0 \cr  } 
\right)$$
$${\bf V}^{g_g\;(-1)}=s^{-1}\;
 \left(
\matrix{ 0 & 0 & 0 & - \frac{1}{4}   & \
-1 \cr 0 & 0 & 0 & 0 & \frac{1}{4} \cr 0 & 0 & 0 & -\frac{s}
   {2} & 0 \cr - \frac{1}{4}   & 0 & -\frac{s}
   {2} & 0 & 0 \cr -1 & \frac{1}{4} & 0 & 0 & 0 \cr  } 
\right)$$
$${\bf V}^{g_g\;(1)}=s^{-1}\;
 \left(
\matrix{ -1 & - \frac{1}{4}   & 0 & 0 & 0 \cr - \frac{1}
     {4}   & 0 & \frac{s}{2} & 0 & 0 \cr 0 & 
\frac{s}{2} & 0 & 0 & 0 \cr 0 & 0 & 0 & 0 & - \frac{1}{4} \
  \cr 0 & 0 & 0 & - \frac{1}{4}   & 1 \cr  } 
\right)$$
$${\bf V}^{g_g\;(2)}=s^{-1}\;
 \left(
\matrix{ 0 & - \frac{1}{4}   & -\frac{s}
   {2} & 0 & 0 \cr - \frac{1}{4}   & -1 & 0 & 0 & 0 \cr 
-\frac{s}{2} & 0 & 0 & 0 & 0 \cr 0 & 0 & 0 & 1 & \frac{1}
   {4} \cr 0 & 0 & 0 & \frac{1}{4} & 0 \cr  } 
\right)$$

$${\bf V}^{h_g\;(-2)}=\frac {\cos\alpha} {\sqrt{20}}
 \left(
\matrix{ 0 & 0 & 0 & s & 0 \cr 0 & 0 & 0 & s & s \cr 0 & 0 & 0 & 0 & -1 \cr s & s & 0 & 0 & 0 \cr 0 & s & -1 & 0 & 0 \cr  } 
\right)
+ \frac {\sin\alpha} {2\; s }
\left(
\matrix{ 0 & 0 & 0 & 1 & 0 \cr 0 & 0 & 0 & -1 & 1 \cr 0 & 0 & 0 & 0 & s \cr 1 & -1 & 0 & 0 & 0 \cr 0 & 1 & s & 0 & 0 \cr  } 
\right)$$
$${\bf V}^{h_g\;(-1)}=\frac {\cos\alpha} {\sqrt{20}}
 \left(
\matrix{ 0 & 0 & 0 & s & -s \cr 0 & 0 & 0 & 0 & -s \cr 0 & 0 & 0 & 1 & 0 \cr s & 0 & 1 & 0 & 0 \cr -s & -s & 0 & 0 & 0 \cr  } 
\right)
+ \frac {\sin\alpha} {2\; s }
\left(
\matrix{ 0 & 0 & 0 & -1 & -1 \cr 0 & 0 & 0 & 0 & 1 \cr 0 & 0 & 0 & s & 0 \cr -1 & 0 & s & 0 & 0 \cr -1 & 1 & 0 & 0 & 0 \cr  } 
\right)$$
$${\bf V}^{h_g\;(0)}=\frac {\cos\alpha} {\sqrt{20}}
 \left(
\matrix{ 1 & 0 & 0 & 0 & 0 \cr 0 & 1 & 0 & 0 & 0 \cr 0 & 0 & \
-4 & 0 & 0 \cr 0 & 0 & 0 & 1 & 0 \cr 0 & 0 & 0 & 0 & 1 \cr  } 
\right)
+ \frac {\sin\alpha} {2 }
\left(
\matrix{ -1 & 0 & 0 & 0 & 0 \cr 0 & 1 & 0 & 0 & 0 \cr 0 & 0 & 0 & 0 & 0 \cr 0 & 0 & 0 & 1 & 0 \cr 0 & 0 & 0 & 0 & -1 \cr  } 
\right)$$
$${\bf V}^{h_g\;(1)}=\frac {\cos\alpha} {\sqrt{20}}
 \left(
\matrix{ -s & s & 0 & 0 & 0 \cr s & 0 & \
-1 & 0 & 0 \cr 0 & -1 & 0 & 0 & 0 \cr 0 & 0 & 0 & 0 & s \cr 0 & 0 & 0 & s & s \cr  } 
\right)
+ \frac {\sin\alpha} {2\; s }
\left(
\matrix{ -1 & -1 & 0 & 0 & 0 \cr -1 & 0 & -s & 0 & 0 \cr 0 & -s & 0 & 0 & 0 \cr 0 & 0 & 0 & 0 & -1 \cr 0 & 0 & 0 & -1 & 1 \cr  } 
\right)$$
$${\bf V}^{h_g\;(2)}=\frac {\cos\alpha} {\sqrt{20}}
 \left(
\matrix{ 0 & s & 1 & 0 & 0 \cr s & -s & 0 & 0 & 0 \cr 1 & 0 & 0 & 0 & 0 \cr 0 & 0 & 0 & s & -s \cr 0 & 0 & 0 & -s & 0 \cr  } 
\right)
+ \frac {\sin\alpha} {2\; s }
\left(
\matrix{ 0 & 1 & -s & 0 & 0 \cr 1 & 1 & 0 & 0 & 0 \cr -s & 0 & 0 & 0 & 0 \cr 0 & 0 & 0 & -1 & -1 \cr 0 & 0 & 0 & -1 & 0 \cr  } 
\right) \ .$$

In the static JT effect, the kinetic term in $P_{\tau i m}^2$ is ignored,
and the problem is to study the five Born-Oppenheimer (BO) potential sheets
given by the sum of each eigenvalue of the electronic problem plus the
harmonic restoring forces.
The $H_u\otimes a_g$ part has a purely diagonal coupling matrix.
As it does not split the electronic degeneracy, it is trivially
separated from the coupling to the other modes and can be treated as a
displaced oscillator.

When a single electron (hole) is placed in the $H_u$ orbital, the molecule
distorts in such a way that the lowest (highest) BO sheet is lowered
(raised) in energy as much as possible.
The coupling to a $g_g$ mode leads to 10 equivalent absolute minima of
$D_3$ local symmetry (Ceulemans and Fowler 1990, Manini and De Los Rios
2000) of the BO potential.
The optimal distortion is realized for $|q_{\rm s}^{g_g}|= g_{g_g}/3$, with
a corresponding potential energy lowering of $g_{g_g}^2 ~
\hbar\omega_{g_g}/18$.
The $r=2$ part of the coupling to the $h_g$ modes (corresponding to the
$\sin\alpha$ terms in ${\bf V}^{h_g\;(m)}$ above) contributes to these same
minima in an equivalent way, with $g_{g_g}$ replaced by
$g_{h_g}\sin\alpha$.
However, the $r=1$ component ($\cos\alpha$ terms in ${\bf V}^{h_g\;(m)}$)
of the coupling favours the six classical stable minima of local $D_5$
symmetry (Ceulemans and Fowler 1990, Manini and De Los Rios 2000).
The optimal distortion at these minima is $|q_{\rm s}^{h_g}|=
g_{h_g}\cos\alpha/\sqrt 5$, for an energy lowering of $g_{h_g}^2 \cos^2
\alpha ~ \hbar\omega / 10$.

The simultaneous linear coupling to several modes will generally lead to a
cumulative distortion and to an energy gain which is the sum of the
individual energy gains.
However, the form of the coupling (\ref{real interaction Hamiltonian})
prevents the molecule to gain energy through both kinds of couplings.
The system shall choose between a $D_3$ and a $D_5$ distortion,
depending which one is energetically more convenient for given specific
values of the couplings, vibration frequencies, and $H_u$ orbital
electronic filling.
The calculation of the following section determines in particular which one
of the two types of distortions prevails in C$_{60}^+$.

\section{Calculation of the couplings and results}
\label{dft:sec}

We compute the molecular electronic structure within the density functional
theory (DFT) in the local density approximation.
The C$_{60}$ molecule is repeated  periodically in a large fcc supercell
lattice.
The conventional supercell side is $a=18.5$~\AA, so that the distance
between the centers of two neighboring copies of the molecule is 13.1~\AA,
suitably much larger than the fully relaxed equilibrium (opposite C-C) ball
diameter, about 7.053~\AA.
Since we aim at describing the single molecule -- and indeed our 
molecules are well isolated -- no sampling of the Brillouin zone is 
called for, and calculations of the charge density are done using
the $k$=0 wavefunctions.
We use ultrasoft pseudopotentials (Vanderbilt 1990) for C (Favot and Dal
Corso 1999).
The plane-waves basis set is cut off at $E_{\rm cut}=27$ Ry (charge
density cutoff = 162 Ry).
Test calculations with higher cutoff or larger cell size $a$ gave
equivalent results.

Based on this electronic structure calculation, we used next density
functional perturbation theory (Baroni {\it et al.} 1987) to compute three
independent rows of the dynamical matrix.
Icosahedral symmetry is then used to recover the full matrix, which
determines the normal modes $\vec\xi_{i,s}$ and frequencies $\omega_i$
(Giannozzi and Baroni 1994) of the molecule.
We obtained frequencies (see Table~\ref{frequencies:tab}) in good agreement
with experiment (Prassides {\it et al.} 1991, Zhou {\it et al.} 1992), as
well as with previous calculations (Giannozzi and Baroni
1994, Negri {\it et al.} 1988, Kohanoff {\it et al.} 1992).

\begin{table}[t]
\begin{center}
\begin{tabular}{lrrr}
\hline\hline
mode            &       Experim.& Giannozzi \& Baroni    & this work\\
\hline
$a_g (1)$       &       496     &       495     &       500     \\
$a_g (2)$       &       1470    &       1504    &       1511    \\
\hline
$h_g (1)$       &       271     &       259     &       261     \\
$h_g (2)$       &       437     &       425     &       429     \\
$h_g (3)$       &       710     &       711     &       718     \\
$h_g (4)$       &       774     &       783     &       784     \\
$h_g (5)$       &       1099    &       1120    &       1119    \\
$h_g (6)$       &       1250    &       1281    &       1275    \\
$h_g (7)$       &       1428    &       1450    &       1456    \\
$h_g (8)$       &       1575    &       1578    &       1588    \\
\hline
$t_{1u} (1)$	&	527	&	527	&	533	\\
$t_{1u} (2)$	&	576	&	586	&	588	\\
$t_{1u} (3)$	&	1183	&	1218	&	1212	\\
$t_{1u} (4)$	&	1428	&	1462	&	1469	\\
\hline\hline
\end{tabular}
\end{center}
\caption{Eigenfrequencies (in cm$^{-1}$) for the Raman- and IR-active
vibrational modes of C$_{60}$ molecule: comparisons with experimental
values (Prassides {\it et al.} 1991, Zhou {\it et al.} 1992, Gunnarsson
{\it et al.} 1995) and previous calculation (Giannozzi and Baroni 1994).
\label{frequencies:tab}}
\end{table}

\begin{figure}[t]
\centerline{
\psfig{file=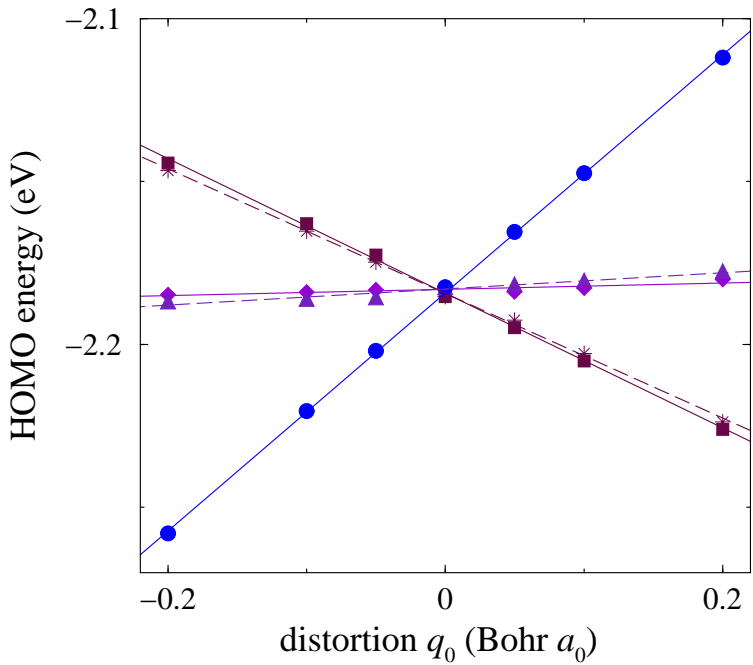,width=10.0cm}
}
\caption{ The splitting of the HOMO degenerate level when the molecule is
distorted according to the 6th mode $h_g$ (1250~cm$^{-1}$).  Points: DFT
data.  Lines: linear fits, giving the coupling parameters $g_{h_g 6}$ and
$\alpha_{h_g 6}$ of Table~\protect\ref{couplings:tab}.
\label{split:fig}}
\end{figure}

To determine the e-v couplings for the linearly coupled modes, we proceed
subsequently to displace the atomic positions from the equilibrium position
along each of the normal modes, choosing a suitable eigenvector combination
in the linear space of each degenerate vibration.
In particular for each $h_g$ mode we selected the $q_0$ displacement,
corresponding to the totally symmetric combination $\vec\xi_{i,0}$ of the
distortions $\vec\xi_{i,s}$ with respect to an (arbitrarily chosen) $D_5$
subgroup of the molecular symmetry group.
The five initially degenerate (really, only nearly degenerate, owing
to a weak cubic splitting due to the artificial supercell lattice) 
$H_u$ Kohn-Sham eigenvalues split
under this distortion with a pattern given by the eigenvalues of 
${\bf V}^{h_g\;(0)}$.
We applied a displacement of the atomic positions along each of the eight
normal-mode unit vectors, $\vec\xi_{i,0}$ with a prefactor ranging from
-0.1 to 0.1~\AA.
In Fig.~\ref{split:fig} we plot as an example the resulting energies 
for the sixth $h_g$ mode.
The pattern generated by ${\bf V}^{h_g\;(0)}$ should be 1+2+2 (a state
separated by two pairs of twofold-degenerate states). The small residual
splittings of these twofold degeneracies, due to the cubic crystal field
and higher-than-linear couplings, give an estimate of the accuracy of the
method.
By standard linear fitting and comparison with Eq.~(\ref{real interaction
Hamiltonian}), we obtained directly the linear dimensionless coupling
coefficients $g_{h_gi}$ $\alpha_{h_gi}$ collected in
Table~\ref{couplings:tab}.
We determined the sign of $\alpha_{h_gi}$ by applying a distortion along
$\vec\xi_{i,1}$, and comparing the splitting of the HOMO with the
eigenvalues of ${\bf V}^{h_g\;(1)}(\pm \alpha_{h_gi})$.
Following the same procedure we derived the couplings for the $g_g$ modes, 
by applying here $q_{-1}$ distortions.
The resulting linear coupling coefficients $g_{g_gi}$ are also collected in
Table~\ref{couplings:tab}.

\begin{table}[t]
\begin{center}
\begin{tabular}{rcccccccr}
\hline
\hline
$\hbar\omega_{\tau i}$	&	coupl.	&
$g_{\tau i}$	&	$\alpha_{\tau i}$	
&	$q_{\rm s}(D_5)$&	$q_{\rm s}(D_3)$	&
$E_{\rm s}(D_5)$	&$E_{\rm s}(D_3)$	&	
$\tilde\lambda$\ \ \ \\
cm$^{-1}$		&	eV/$a_0$	&    &deg
&	$a_0$	&	$a_0$ &
meV	&	meV &
meV	\\
\hline
$a_g$\\
500     &       0.03    &       0.059   &       -       &       0.00    &
0.00    &       0.0     &       0.0     &             0.1     \\
1511    &       0.63    &       0.274   &       -       &       0.01    &
0.01    &       1.8     &       1.8     &             3.5     \\
\hline
$g_g$	\\
483     &       0.31    &       0.757   &       -       &       -    &
0.04    &       0.0     &       1.9     &             8.6     \\
567     &       0.05    &       0.102   &       -       &       -    &
0.00    &       0.0     &       0.0     &             0.2     \\
772     &       0.67    &       0.800   &       -       &       -    &
0.03    &       0.0     &       3.4     &            15.3    \\
1111    &       0.90    &       0.624   &       -       &       -    &
0.02    &       0.0     &       3.0     &            13.4    \\
1322    &       0.43    &       0.228   &       -       &       -    &
0.01    &       0.0     &       0.5     &             2.1     \\
1519    &       1.08    &       0.467   &       -       &       -    &
0.01    &       0.0     &       2.3     &            10.3    \\
\hline
$h_g$	\\
261     &       0.50    &       3.042   &        -0.1   &       0.27    &
0.00    &       30.0    &       0.0     &            75.0    \\
429     &       0.43    &       1.223   &       30.1    &       0.07    &
0.03    &       6.0     &       1.1     &            19.9    \\
718     &       0.75    &       0.995   &       89.4    &       0.00    &
0.04    &       0.0     &       4.9     &            22.0    \\
785     &       0.67    &       0.784   &        -2.3   &       0.04    &
0.00    &       6.0     &       0.0     &            15.0    \\
1119    &       0.32    &       0.221   &       76.6    &       0.00    &
0.01    &       0.0     &       0.4     &             1.7     \\
1275    &       0.93    &       0.519   &       28.0    &       0.02    &
0.01    &       3.3     &       0.5     &            10.7    \\
1456    &       2.09    &       0.962   &       28.1    &       0.03    &
0.01    &       13.0    &       2.1     &            41.7    \\
1588    &       2.15    &       0.869   &      -31.1    &       0.03    &
0.01    &       10.9    &       2.2     &            37.1    \\
\hline
\hline
\end{tabular}
\end{center}
\caption{Computed mode eigenfrequencies and e-v linear coupling parameters
of the $H_u$ HOMO in C$_{60}$.  The JT distortion magnitudes $q_{\rm s}$
and the classical stabilization energies $E_{\rm s}$ are tabulated
for both $D_5$ 
and $D_3$ distortions and for one hole in the HOMO. 
The largest total JT energy gain is realized by the $D_5$ distortions.
We also show the contribution of each mode to the resistivity and
superconductivity total coupling $\tilde\lambda=\lambda/N_1(0)$ defined in
Eq.~(\ref{M:lambdatransport})
[$N_1(0)=$ density of states per spin {\em per band} at the Fermi level].
\label{couplings:tab}}
\end{table}

For convenience, we also report in Table~\ref{couplings:tab} the amount of
optimal JT distortion pertinent to each mode when the HOMO level is
occupied by one electron/hole, and the corresponding energy lowering
$E_{\rm s}$ for both $D_5$ and $D_3$ minima.
Note in particular the large coupling associated to the lowest $h_g$
mode, the corresponding distortion leading to an energy lowering 
practically equal to its quantum
$\hbar\omega$.

Adding up the JT energy gain of the individual modes, we estimate the total
classical potential energy lowering in C$_{60}$.  The $D_5$ minima gain
$E_{\rm s}$=71~meV, while the $D_3$ minima gain only $E_{\rm s}$=22~meV
(the contribution of the $a_g$ modes -- 2~meV -- being included in both
cases).  It is therefore apparent that the C$_{60}^{+}$ ion will choose, at
least within linear coupling, the $D_5$ distortion.
As was shown in (Manini and De Los Rios 2000), the possibility of a switch
to a nondegenerate $A_u$ dynamical JT GS occurs, for large coupling
strength $g\geq 6$, only under the condition that the $D_3$ minima are
energetically lower or equal to the $D_5$ minima.
This settles finally the issue of the dynamical JT GS symmetry of this
molecular ion: it is a regular Berry-phase vibronic state of symmetry
$H_u$, like the parent electronic state (Manini and De Los Rios 2000).  
No level crossing to a nondegenerate $A_u$ state is predicted to occur for
C$_{60}^+$.

\begin{figure}[t]
\centerline{
\psfig{file=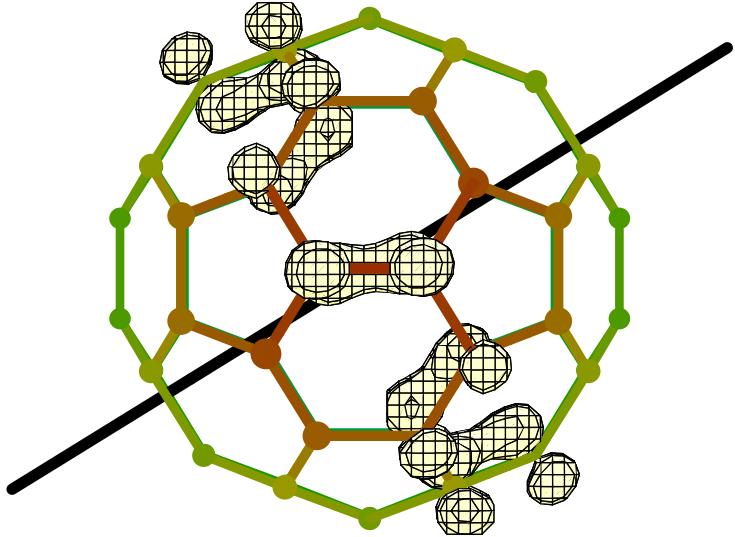,width=10.0cm}
}
\caption{ Electronic charge density distribution of the hole in C$_{60}^+$
at the $D_5$ JT minimum along a $q_0$ distortion of any $h_g$ mode.
Out of the six equivalent ones, this particular minimum is defined by the
$D_5$ axis drawn as a bold ``spit'' piercing the ball through two
pentagons.
The contours are drawn at 30\% of the maximum density.
\label{bandHu5:fig}}
\end{figure}

A static JT coupling resolves the degeneracy of the $H_u$ level: in a distorted
configuration it is possible to distinguish individual levels within the
HOMO, energy-wise.
Figure~\ref{bandHu5:fig} depicts the square modulus of the hole
wavefunction for a $D_5$ minimum.
The probability density appears to be localized on an equatorial conjugated
band, where the poles are the opposite pentagons centered around the $D_5$
axis we chose among the six possible ones.

As a check, with the same method used above to calculate the hole-vibration
couplings of the HOMO (Fig.~\ref{split:fig}), we also computed the
electron-vibration couplings of the LUMO and obtained the values in
Table~\ref{minus:tab}.
The total static JT potential energy lowering is 41~meV, of which 
3~meV due to the $a_g$ modes, and 38~meV due to the $h_g$ modes.
These values are generally in line with those calculated by previous
authors (Antropov {\it et al.} 1993, Varma {\it et al.} 1991, Lannoo {\it
et al.}  1991), although there are some differences in the details. Error
bars in theoretical determinations of JT couplings of fullerene have proven
surprisingly large, possibly reflecting and amplifying errors in the
vibrational eigenvectors.
Of course as is well known, somewhat larger energy gains are obtained when
the true dynamical JT problem, including a full quantum treatment of the
vibrons is considered (Auerbach {\it et al.}  1994, Manini {\it et al.}
1994).
We shall leave this calculation in C$_{60}^{n+}$ for future work.

\begin{table}[t]
\begin{center}
\begin{tabular}{rccccr}
\hline
\hline
$\hbar\omega_{\tau i}$	&	coupl.	&
$g_{\tau i}$	&	$q_{\rm s}$     &
$E_{\rm s}$	&
	$\tilde\lambda$\ \ \ \\
cm$^{-1}$		&	eV/$a_0$	&
&	$a_0$	&
meV	&	meV	\\
\hline
$a_g$\\
500     &       0.07    &       0.157   &       0.01    &       0.2     &
      0.4     \\
1511    &       0.78    &       0.340   &       0.01    &       2.7     &
      5.4     \\
\hline
$h_g$	\\
261     &       0.07    &       0.412   &       0.08    &       2.7     &
     13.7    \\
429     &       0.17    &       0.489   &       0.07    &       6.3     &
     31.7    \\
718     &       0.26    &       0.350   &       0.04    &       5.5     &
     27.3    \\
785     &       0.19    &       0.224   &       0.03    &       2.4     &
     12.2    \\
1119    &       0.28    &       0.193   &       0.02    &       2.6     &
     12.9    \\
1275    &       0.25    &       0.138   &       0.01    &       1.5     &
      7.6     \\
1456    &       0.69    &       0.315   &       0.03    &       9.0     &
     44.8    \\
1588    &       0.72    &       0.289   &       0.02    &       8.2     &
     41.2    \\
\hline
\hline
\end{tabular}
\end{center}
\caption{Computed mode eigenfrequencies and e-v linear coupling parameters
for the $T_{1u}$ LUMO of C$_{60}$.  For each mode, we report The JT
distortion magnitudes $q_{\rm s}$, the classical stabilization energies
$E_{\rm s}$, and the contribution to $\tilde\lambda=\lambda/N_1[0]$.
\label{minus:tab}}
\end{table}

\begin{figure}[t]
\centerline{
\psfig{file=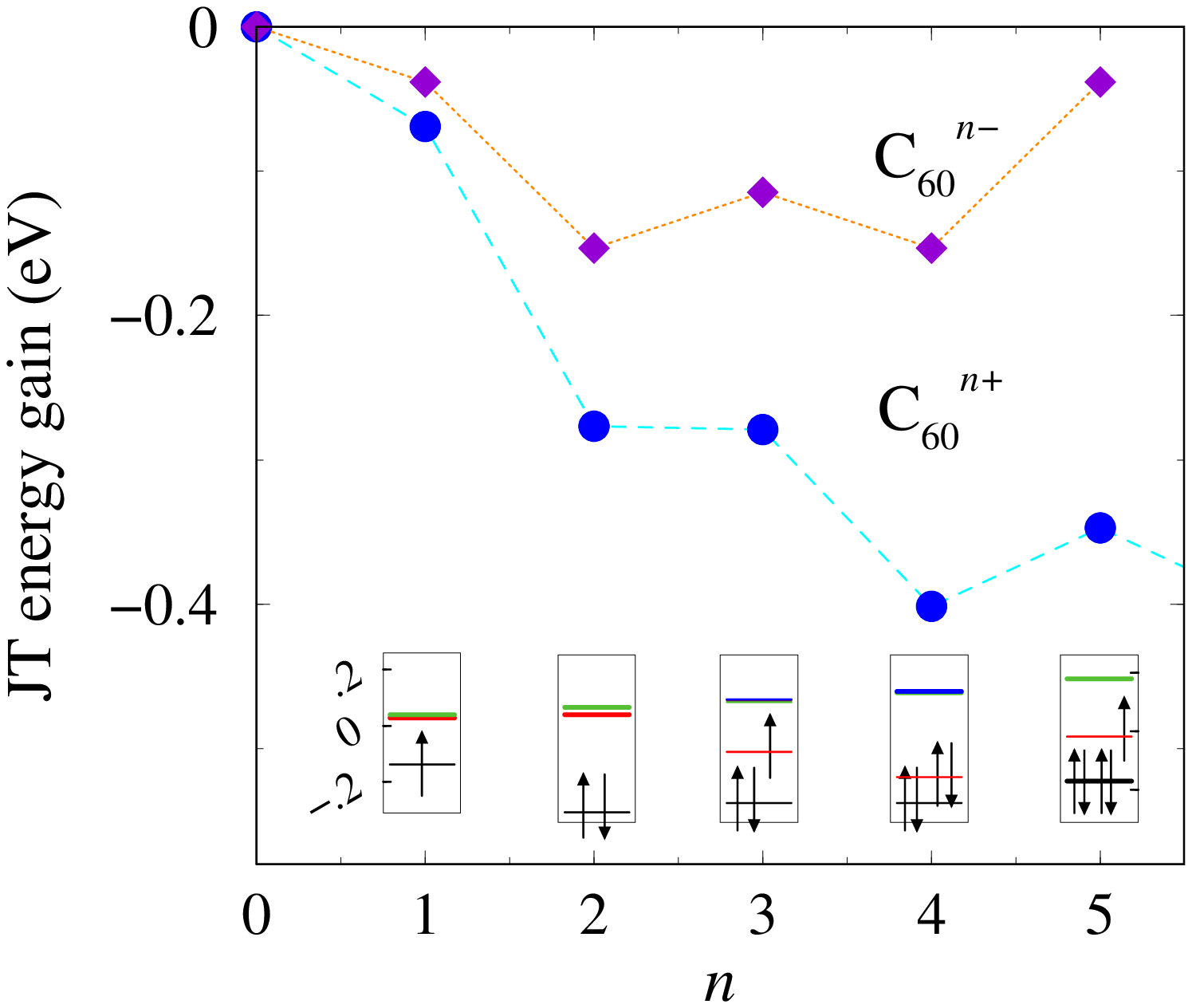,width=10.0cm}
}
\caption{ The static JT energy gain (excluding the $a_g$ modes
contribution) of C$_{60}^{n+}$, as a function of the filling $n$ of the
$H_u$ molecular level.
Electron-hole symmetrical values are obtained beyond half-filling $n=5$.
The corresponding energies for the $T_{1u}$ LUMO (C$_{60}^{n-}$) are also
plotted for comparison.
Note the much larger gains (up to a factor 3) in the $H_u$ case.
For each $n$, the insets show the corresponding split HOMO electronic
configuration (all in the same scale of eV).
\label{eclass_n:fig}}
\end{figure}

When more than one electron/hole occupies the HOMO of C$_{60}$, larger
distortions and more important energy gains are expected.
In particular, the $a_g$ contribution for C$_{60}^{n+}$ is simply:
\begin{equation}
E_{\rm s}^{a_g}(n)=n^2 E_{\rm s}^{a_g}(1),
\label{ag_energy}
\end{equation}
which can become as large as 179~meV for $n=10$ holes.
Consider now the nontrivial JT part of the coupling, that to $g_g$ and
$h_g$ modes.
For $n=2$ electrons/holes (spin singlet configuration) in the HOMO orbital,
the distortions simply become twice as large as in the $n=1$ case, with JT
energy gains which are four times larger.
Even though $n=1,2$ electrons/holes take advantage only of the $D_5$
stabilization energy, additional electrons/holes can benefit from the extra
HOMO splitting induced by the $h_g$($r=2$) plus $g_g$ coupling.
To test this, we relaxed completely the molecular structure in the
64-dimensional space of the 8 $h_g$ and $g_g$ modes, and determined the
minima of the total potential energy, filling the five BO sheets as drawn
in the insets of Fig.~\ref{eclass_n:fig}.
For simplicity we only considered at this stage low-spin configurations, 
generally the most favored by JT.
The resulting energy gains $E_{\rm s}^{g_g+h_g}(n)$ are reported in
Fig.~\ref{eclass_n:fig}.
The energy lowering is maximum for $n=4$ (and $n=6$): it is as large as
401~meV, 
 compared to a modest $E_{\rm s}^{h_g}(n=2,4)=153$~meV in C$_{60}^-$.
Similarly to what happens in the $T_{1u}$ LUMO case (see
Fig.~\ref{eclass_n:fig} and Manini {\it et al.} 1994), the half-filled
configuration $n=5$ is slightly unfavorable (by JT energetics) with respect
to the neighboring $n=4$ and $n=6$ states.
We find that the contributions of the $g_g$ modes, strictly zero for
$n=1,2$, are small ($\sim 1$~meV) but nonzero in the $3\leq n\leq 7$
configurations.
This indicates that for such large fillings the many-modes JT system
(Manini and Tosatti 1998) can take some advantage also of the ``losing''
$g_g$ + $h_g(r=2)$ part of the coupling (favoring $D_3$ minima for 
$n=1,2$).
The insets of Fig.~\ref{eclass_n:fig}, indicate that the largest
displacement of a single level in the $H_u$ HOMO,
about 0.28~eV, 
is realized for $n=2$.
Finally, we have particle-hole symmetry $E_{\rm s}^{g_g+h_g}(n)=E_{\rm
s}^{g_g+h_g}(10-n)$, at an opposite minimum distortion $\vec q_{\rm
s}(n)=-\vec q_{\rm s}(10-n)$.
For the special case $n=5$ this means that the configuration drawn in the
inset and the one obtained reflecting the 5 levels through zero give both
the same optimal energy (at opposite distortions).

Application of the above results to real C$_{60}^{n+}$ must await the
inclusion of electron-electron Coulomb repulsion.
Coulomb interactions will generally compete with JT coupling and favor
high-spin configurations, which, in turn, are generally less favorable for
JT.
(For example for $n=2$ the triplet configurations has a JT gain of
100~meV 
instead of 277~meV for the singlet.)
We shall return to a more detailed description of C$_{60}^{n+}$ ions in 
later work.

\section{Solid-State Transport}
\label{solid:sec}

In order to relate the previously calculated electron-vibron coupling 
constants to relevant solid state physical quantities, we start by  
re-deriving its contribution to the transport relaxation time.
We are interested in particular to the features of an orbitally 
degenerate band as in the charged C$_{60}$ case.
The Boltzmann equation in the presence of a uniform and static electric
field $\vec{E}$ reads
\begin{equation}
e\vec{v}_{k\mu}\cdot \vec{E} \;
\frac{\partial n^0(\epsilon_{k\mu})}{\partial \epsilon_{k\mu}} 
=   \left(\frac{\partial n_{k,\mu}}{\partial t}\right)_{coll}.
\label{M:Boltzmann}
\end{equation}
being $n_{k,\mu}$ and $\epsilon_{k\mu}$ the occupation number and energy 
at momentum $k$ for orbital $\mu$, and  
$\vec{v}_{k\mu}=\partial \epsilon_{k\mu}/\partial\vec{k}$. 

Within the relaxation time approximation, $\partial
n_{k,\mu}/\partial t = \delta n_{k,\mu}/\tau_{k\mu}$
so that the variation with respect to equilibrium becomes
\begin{equation}
\delta n_{k,\mu} = - 
\frac{\partial n^0(\epsilon_{k\mu})}{\partial \epsilon_{k\mu}} 
e \vec{v}_{k\mu}\cdot \vec{E} \tau_{k\mu}, 
\label{M:relaxation time approximation}
\end{equation}
and the conductivity is obtained in the form
\begin{equation}
\sigma = -e^2\frac{1}{3V}\sum_{k\mu} \tau_{k\mu} 
\vec{v}_{k\mu}  \cdot \vec{v}_{k\mu}   
\frac{\partial n^0(\epsilon_{k\mu})}{\partial \epsilon_{k\mu}},
\end{equation} 
$V$ being the volume, measured in cell units. 

As is usual in the treatment of different scattering mechanisms 
within Fermi's golden rule approximation, 
one has to sum the inverse of the corresponding relaxation times. 
The collision term due to the electron-vibron 
coupling is given, within Fermi's golden rule, by 
\begin{eqnarray}
&&\left(\frac{\partial n_{k,\mu}}{\partial t}\right)_{coll} 
= \frac{2\pi}{\hbar}\frac{1}{2 V} \sum_q \sum_{i,\tau,m,\nu}
\left(\frac{g_{\tau i}\hbar\omega_{\tau i}}{2}\right)^2 
\left| V^{\tau(m)}_{\nu\mu}(\alpha_{\tau i})\right|^2 
\left\{\phantom{\Sigma_a^b} \right . \nonumber \\
&& n_{k-q,\nu}\left(1-n_{k,\mu}\right)
\left[N^{\tau i}_{q,m}\delta\left(\epsilon_{k\mu}
-\epsilon_{k-q\nu} - \hbar\omega_{\tau i}\right)
+ \left(N^{\tau i}_{-q,m}+1\right)\delta\left(\epsilon_{k\mu}
-\epsilon_{k-q\nu} + \hbar\omega_{\tau i}\right)\right] \nonumber \\
&& \left . - n_{k,\mu}\left(1-n_{k-q,\nu}\right)
\left[N^{\tau i}_{-q,m}\delta\left(\epsilon_{k-q\nu}
-\epsilon_{k\mu} - \hbar\omega_{\tau i}\right)
+ \left(N^{\tau i}_{q,m}+1\right)\delta\left(\epsilon_{k-q\nu}
-\epsilon_{k\mu} + \hbar\omega_{\tau i}\right)\right]\right\}  \ . 
\nonumber \\
\label{M:collision term}
\end{eqnarray}
Here $N^{\tau i}_{q,m}$ is the Bose-Einstein occupation 
number of phonons of crystal momentum
$q$, symmetry $\tau = a_g,g_g,h_g$, mode $i$, component $m$.
$n_{k,\mu}$ is the Fermi occupation of HOMO band component $\mu=-2,\dots,2$
at the given temperature and chemical potential.
At equilibrium, the collision term is zero. We expand to first order in the
deviation from equilibrium.

First, let us consider the case $T\gg {\rm Max}_i(\hbar\omega_i)$ 
but still $T\ll T_F$ (the Fermi temperature). In this case 
$$
N^{\tau i}_{q,m} \simeq \frac{k_{\rm B} T}{\hbar\omega_{\tau i}} \gg 1,
$$
and 
\begin{eqnarray}
&&\left(\frac{\partial n_{k,\mu}}{\partial t}\right)_{coll} 
= - \frac{2\pi}{\hbar}\frac{1}{2 V} \sum_{p} \sum_{i,\tau,m,\nu}
\left(\frac{g_{\tau i}\hbar\omega_{\tau i}}{2}\right)^2 
\left| V^{\tau(m)}_{\nu\mu}(\alpha_{\tau i})\right|^2 
\frac{k_{\rm B} T}{\hbar\omega_{\tau i}}
\left\{\phantom{\Sigma_a^b} \right . \nonumber \\
&& \;\left. \left(\delta n_{k,\mu}-\delta n_{p,\nu}\right) 
\left[ \delta\left(\epsilon_{k\mu}
-\epsilon_{p\nu} - \hbar\omega_{\tau i}\right) 
+ \delta\left(\epsilon_{k\mu}
-\epsilon_{p\nu} + \hbar\omega_{\tau i}\right)\right]\right\}.
\label{M:collision term final}
\end{eqnarray}

Since $T\gg {\rm Max}_i(\hbar\omega_i)$, after inserting 
(\ref{M:relaxation time approximation}) into  
(\ref{M:collision term final}) and using the Boltzmann equation, 
we find that  
\begin{eqnarray}
\frac{1}{\tau_{k\mu}} &=& 
\frac{2\pi}{\hbar}\frac{1}{2 V} \sum_{p} \sum_{i,\tau,m,\nu}
\left(\frac{g_{\tau i}\hbar\omega_{\tau i}}{2}\right)^2 
\left| V^{\tau(m)}_{\nu\mu}(\alpha_{\tau i})\right|^2 
\frac{k_{\rm B} T}{\hbar\omega_{\tau i}}
\left\{\phantom{\Sigma_a^b} \right . \nonumber \\
&& \;\left. 
\left(1-\frac{\vec{v}_{p\nu}\cdot\vec{n}}
{\vec{v}_{k\mu}\cdot\vec{n}}\right)
\left[ \delta\left(\epsilon_{k\mu}
-\epsilon_{p\nu} - \hbar\omega_{\tau i}\right) 
+ \delta\left(\epsilon_{k\mu}
-\epsilon_{p\nu} + \hbar\omega_{\tau i}\right)\right]\right\},
\end{eqnarray}
where $\vec{n}$ is a unit vector parallel to the electric field. 
 
If we approximately take $\epsilon_{k\mu}=\epsilon_k$ we find, since the 
electron-vibron matrices are symmetrical, 
\begin{eqnarray}
\frac{1}{\tau_{k}} &=& 
\frac{\pi}{\hbar} \sum_{i,\tau,m,}
\left(\frac{g_{\tau i}\hbar\omega_{\tau i}}{2}\right)^2 
~ \frac{{\rm Tr}\left[\left({\bf V^\tau}\right)^2\right]}d 
~ \frac{k_{\rm B} T}{\hbar\omega_{\tau i}}
\left\{\phantom{\Sigma_a^b} \right . \nonumber \\
&& \;\left. \frac{1}{V}\sum_p
\left(1-\frac{\vec{v}_{p}\cdot\vec{n}}
{\vec{v}_{k}\cdot\vec{n}}\right)
\left[ \delta\left(\epsilon_{k}
-\epsilon_{p} - \hbar\omega_{\tau i}\right) 
+ \delta\left(\epsilon_{k}
-\epsilon_{p} + \hbar\omega_{\tau i}\right)\right]\right\}\ .
\end{eqnarray}
where $d$ is the orbital multiplicity, 5 for the HOMO-derived band of
C$_{60}$.
Therefore, the relaxation rate in the linear $T$ regime 
turns out to be roughly proportional to 
\begin{equation}
\tilde\lambda = \sum_{i,\tau,m}
\frac{g_{\tau i}^2 \hbar\omega_{\tau i}}{4} 
~ \frac{{\rm Tr} \left[\left({\bf V^\tau}\right)^2\right]}d 
\ ,
\label{M:lambdatransport}
\end{equation} 
related to the conventional dimensionless ep-phonon coupling $\lambda$ by
\begin{equation}
\lambda =  N_1(\epsilon_F) ~ \tilde\lambda
\ ,
\label{lambdadef:eq}
\end{equation} 
where $N_1(\epsilon_F)$ is the density of states per spin {\em per band} at
the Fermi energy.

By using Eq.~(\ref{M:lambdatransport}) and the calculated e-v coupling
parameters of Table~\ref{couplings:tab}, we find for holes in the $H_u$
HOMO-derived band
\begin{equation}
\tilde \lambda_+ \simeq 0.277~{\rm eV}.  
\label{M:lambda+}
\end{equation}
This value can be compared with that calculated similarly, using
our couplings of Table~\ref{minus:tab} for electrons in the $T_{1u}$ band
\begin{equation}
\tilde\lambda_-  \simeq 0.197~{\rm  eV}.
\label{M:lambda-}   
\end{equation}
We conclude therefore that holes have a stronger scattering with
intra-molecular optical phonons than electrons, by about a factor 1.4.

To confirm the accuracy of the values computed for the HOMO case, it is
useful to compare those for the LUMO with previous similar calculations. It
is then necessary to take into account the factor $d=3$ which is usually
incorporated in the {\em total} density of states (Lannoo {\it et al.}
1991).
Our $\tilde\lambda/d=66$~meV compares well with previous similar
calculations for the C$_{60}$ LUMO (Antropov {\it et al.} 1993, Varma {\it
et al.} 1991, Lannoo {\it et al.}  1991), ranging from 49 to 68~meV.
On the other hand, empirical estimates based on a fit to 
photoemission data tend to give a value as large as 147~meV
(Gunnarsson {\it et al.} 1995).
The origin of this discrepancy is presently unclear.

\section{Superconductivity}
\label{superc:sec}

The dimensionless electron-vibron coupling which governs the transport
properties, $\lambda$ of Eqs.~(\ref{M:lambdatransport},\ref{lambdadef:eq}),
is not in point of principle coincident with the parameter $\lambda$
determining the superconducting properties. The latter must be determined
by solving the Migdal-Eliashberg equation with the retarded interaction
mediated by the vibrons plus the electron-electron Coulomb repulsion.
However, a simple estimate of the order of magnitude of $\lambda$ can be
obtained by taking the unretarded limit, and imposing a Debye cutoff to the
electron energies.
We assume the electronic band operators to be related to the molecular
creation and annihilation operators through the unitary (orthogonal)
transformation
$$
c_{nk\sigma} = U^{-1}_{n\mu}(k)\, d_{\mu k \sigma},
$$
$n$ being the band index, $\sigma$ the spin.  Let us define a matrix ${\bf
W}$ by

\begin{equation}
W_{\mu\nu} \equiv \sum_{\tau,i,\widehat m}
\frac{g_{\tau i}^2 \hbar\omega_{\tau i}}{4} 
\left(V^{\tau(\widehat m)}_{\mu\nu}\right)^2
= \sum_\tau W^\tau_{\mu\nu} \sum_{i=1}^{n_m(\tau)} 
\frac{g_{\tau i}^2 \hbar\omega_{\tau i}}{4},
\label{M:W}
\end{equation}
as well as 
\begin{equation}
{\cal W}_{nk,mp} \equiv \sum_{\tau,i,\widehat m} \sum_{\mu_1,\mu_2,\mu_3,\mu_4}
\frac{g_{\tau i}^2 \hbar\omega_{\tau i}}{4}
U^{-1}_{n\mu_1}(k)\, V^{\tau(\widehat m)}_{\mu_1\mu_2}\, U_{\mu_2 m}(-p)\,
U_{\mu_3 m}(p)\, V^{\tau(\widehat m)}_{\mu_4\mu_3}\, U^{-1}_{n\mu_4}(-k),
\label{M:Wbis}
\end{equation}

The Bardeen-Cooper-Schrieffer (BCS) gap equation for 
$$
\Delta_{k,n} = \langle 
c^\dagger_{k,\uparrow,n}c^\dagger_{-k,\downarrow,n} \rangle,
$$
reads
\begin{equation}
\Delta_{k,n} = \frac{1}{2V}\sum_{p,m} {\cal W}_{nk,mp} 
\frac{\Delta_{p,m}}{E_{p,m}}
\tanh\left(\beta\frac{E_{p,m}}{2}\right),
\label{M:BCSeq}
\end{equation}
where
$$
E_{p,m}=\sqrt{\left(\epsilon_{pm}-\mu_0\right)^2 + \Delta_{p,m}^2},
$$
$\mu_0$ being the chemical potential.  
The critical temperature is obtained by solving the eigenvalue equation
$$
\delta_{nm}\delta_{kp} - \frac{1}{2V}\sum_{p,m} {\cal W}_{nk,mp}
\frac{1}{\left|\epsilon_{pm}-\mu_0\right|}
\tanh\left(\beta_c\frac{\left|\epsilon_{pm}-\mu_0\right|}{2}\right) = 0.
$$
In general, the BCS gap equation (\ref{M:BCSeq}) leads to interference
between the Cooper pairs belonging to different bands. That, in turn,
increases the critical temperature relative to a situation in which the
pairs do not interfere.  We may therefore foresee two opposite limits of
strongly interfering and of non interfering pairs which, respectively, over
and underestimate the effective coupling strength $\lambda$.

If we assume that averages over the Fermi surface do not depend on the band
indices (interfering pairs, corresponding, for example, to the choice of
Lannoo {\it et al.} 1991), then we can replace ${\cal W}$ with
${\bf W}$, and we find that the critical temperature is determined by the
{\em maximum eigenvalue} of the matrix $\bf W$ in Eq.~(\ref{M:W}).
Under this assumption, the superconducting $\lambda$ is determined
through
\begin{equation}
\lambda = N_1(\epsilon_F){\rm Max}_{\rm eigenvalue}\left({\bf W}\right),
\label{M:lambda}
\end{equation} 

For our $H \otimes (a+g+h)$ e-ph coupling the matrices ${\bf W}^\tau$ are:
\begin{equation}
        {\bf W}^{g_g}=\frac 1{24}
  \pmatrix{
    8 & 1 & 6 & 1 & 8 \cr 1 & 8 & 6 & 8 & 1 \cr 6 & 6 & 0 & 6 & 6 \cr 1 & 8 &
    6 & 8 & 1 \cr 8 & 1 & 6 & 1 & 8 \cr  } 
\end{equation}
\begin{eqnarray}
        {\bf W}^{h_g}&=&\frac 1{60}
  \pmatrix{
    16 & 14 & 9 & 14 & 7 \cr 14 & 16 & 9 & 7 & 14 \cr 9 & 9 & 24 & 9 & 9  \cr
    14 & 7 & 9 & 16 & 14 \cr 7 & 14 & 9 & 14 & 16 \cr  }
 \nonumber\\
&&+\frac{\cos(2 \alpha)}{30}
  \pmatrix{ -2 & 2 & -3 & 2 & 1 \cr 2 & -2 & -3 & 1 & 2 \cr -3 & -3 & 12 & 
     -3 & -3 \cr 2 & 1 & -3 & -2 & 2 \cr 1 & 2 & -3 & 2 & -2 \cr  } 
 \nonumber\\
&&+\frac{\sin(2 \alpha)}{4 \sqrt 5}
  \pmatrix{ 0 & 0 & -1 & 0 & 1 \cr 0 & 0 & 1 & -1 & 0 \cr -1 & 1 & 0 & 1 & 
     -1 \cr 0 & -1 & 1 & 0 & 0 \cr 1 & 0 & -1 & 0 & 0 \cr  } 
\ ,
\end{eqnarray}
while ${\bf W}^{a_g}$ is trivially the unit matrix.
We note that, even though the matrices ${\bf W}^\tau$, for different
$\tau$'s, do not commute, the eigenvector $(1,1,1,1,1)/\sqrt 5$ is an
eigenstate of each of them.
Moreover, for each $\tau$ this eigenstate provides the largest eigenvalue
\begin{equation}
{\rm Max}_{\rm eigenvalue}\left({\bf W}^\tau\right)=\sum_n W^\tau_{mn}=1
\end{equation}
(for any $m$ and $\tau$).
This means that the totally-symmetric paired state, delocalized over all
the five $H_u$ orbitals or bands, is favored by the couplings to all modes.
It also implies that, contrary to molecular JT, the couplings to all modes
cooperate evenly to this superconducting state, and contribute additively to
$\lambda$.
Totally equivalent (even if at first sight apparently different) results
were derived for the $d=3$ case (K$_3$C$_{60}$) in Refs. (Lannoo {\it et
al.}  1991, Rice {\it et al.}  1991).
We note, however, that the claim that orbital degeneracy enhances the
superconducting $\lambda$ through a factor $d$ (Rice {\it et al.}  1991) is
not really justified, as one must at the same time reduce the density of
states from total to single-band, a factor $1/d$ smaller.
We also note that
\begin{equation}
1= 
\sum_n W^\tau_{mn}=
\frac {\sum_{mn} W^\tau_{mn}}d=
\frac {\sum_{mn} V^\tau_{mn}V^\tau_{nm}}d=
\frac {\sum_{m} \left[({\bf V}^\tau)^2\right]_{mm}}d =
\frac {{\rm Tr}\left[({\bf V}^\tau)^2\right]}d \ ,
\end{equation}
which shows the identity of the $\tilde\lambda$ computed for
superconductivity to the one obtained for transport in
Eq.~(\ref{M:lambdatransport}).

If, in the opposite limit, the pairs did not interfere between different
bands, the effective $\lambda$ would be reduced by a factor $d$.
Although we cannot identify a physical situation corresponding to this
limit, we can assume that a general case will be intermediate between the
limits (interfering/not-interfering pairs).
For simplicity, we will stick here to the interfering limit.

In summary, is there any enhancement of superconductivity due to orbital
degeneracy? We can still identify one possible source for that, namely
Coulomb pseudopotential.
In fact, we note that, although a large $\lambda$ due to tunneling of the
Cooper pairs between different orbitals/bands, can be seen as orbital
degeneracy enhancing the effective $\lambda$ to the highest eigenvalue of
$\bf W$, there is no corresponding enhancement of the repulsive Coulomb
pseudo-potential $\mu^*$, at least within the Migdal-Eliashberg theory.
The reason is that the main contribution to the Coulomb pseudo-potential is
a charge-charge repulsion which does not include tunneling processes
between different bands, and being band-diagonal it does not get enhanced.
In conclusion, in the above restricted sense, orbital degeneracy may in
principle favor superconductivity.

\section{ Discussion}
\label{discussion:sec}

We have presented a density functional calculation of the linear coupling
of holes/el\-ec\-trons in the $H_u$ and $T_{1u}$ orbitals of the fullerene
molecule to the intra-molecular vibrations.
The coupling to holes is strongest for the $h_g$ modes, and among those to
the lowest-frequency mode $h_g(1)$ around 270 cm$^{-1}$.
The linear static Jahn Teller distortion predicted for C$_{60}^{+}$ by these
couplings corresponds to a $D_5$ distortion, with an energy gain of 71~meV.

The corresponding dynamical JT state expected with the calculated coupling
parameters is a regular Berry-phase vibronic state of symmetry $H_u$, like
the parent electronic state (Manini and De Los Rios 2000).
There is no level crossing to a nondegenerate $A_u$ state, as would have
hypothetically been possible on pure symmetry grounds, had the $D_3$ minima
been the stable ones (Manini and De Los Rios 2000).

In order to connect with important solid state properties including
transport and superconductivity we have formulated a theory of the
Boltzmann relaxation time, and of the BCS-type pairing, suitable for an
orbitally degenerate multiband case with Jahn Teller coupling.
This confirms that the same parameter $\lambda$ determines both transport
and superconducting properties of the multiband degenerate solid.
As we previously observed, not all the computed $g_i$
(Table~\ref{couplings:tab}) are small parameters, and thus weak-coupling
BCS theory is strictly not applicable for C$_{60}^{n+}$.
However, the overall $\lambda$ is still moderate.
Our calculation neglects couplings to acoustic phonons and librations,
which in principle should also contribute to e-ph
scattering.
In addition, similarly to Lannoo {\it et al.}'s calculation (Lannoo {\it et
al.}  1991), we assume that the dispersion of the HOMO band and of the
optical phonons has a negligible effect on the integrated value of
$\lambda$.
(This assumption was tested and proved correct in Ref. (Antropov {\it et
al.} 1993) for the couplings to the LUMO band.)
With all these approximations, Eq.~(\ref{M:lambda}) should provide a
semi-quantitative estimate of the total e-ph scattering.

Assuming conservatively a total average density of states of $\sim
10/0.6~{\rm eV}\approx17$~stat\-es~eV$^{-1}$ for the HOMO band (i.e. a
single-band density of states $N_1(0)$ = 1.7~stat\-es~eV$^{-1}$ per band per
spin), our calculated effective dimensionless $\lambda_+$ for hole
superconductivity in C$_{60}$ is in conclusion about
$\lambda_+\approx0.47$.
The Coulomb pseudopotential $\mu^*$ is not available yet, but possibly in
the same range of values as for negative C$_{60}$ [$\mu^*\sim 0.2\div 0.3$
(Gunnarsson {\it et al.} 1995, Gunnarsson and Zwicknagl 1992)].
With this value of $\lambda_+$, weak coupling would predict
$T_c \sim 1.14~\hbar\omega_{\rm D}\,k_{\rm B}^{-1}
\exp[-1/(\lambda_+ -\mu^*)]\sim 40$~K
for
$\mu^*= 0.2$, and $T_c\sim 5$~K for $\mu^*= 0.3$ (assuming a typical
phonon energy $\omega_D$ of about 1500~K).
This seems of the correct order of magnitude, although somewhat on the low
side, in comparison with $T_c$= 52 K found experimentally.
However, it is difficult to justify weak coupling in this case.

The corresponding value $\lambda_-$ which we obtain for electrons in the
$T_{1u}$ orbitals is, assuming the same bandwidth of 0.6~eV, thus again
$N_1(0) = 1.7$~stat\-es~eV$^{-1}$ per band per spin for the $T_{1u}$ band, 
$\lambda_-\approx 0.33$.
The factor $\lambda_+/\lambda_-=1.4$ of holes relative to electrons is in
qualitative agreement with a larger $T_c$ of the former.
Assuming the same typical phonon frequency, BCS would predict here
$T_c\sim 0.8$~K for $\mu^*= 0.2$, and $T_c\sim 0$~K for $\mu^*= 0.3$.
That is obviously way smaller than the observed superconducting $T_c$=
10 K found experimentally in the field emission transistor (FET) experiment,
let alone the higher values found in the fullerides.

Coming to transport, the measured resistivities for holes are
larger than those of electrons, and this also agrees with a larger 
$\lambda$ value. Calculation of the $T$-linear high temperature
resistivity
$$
\rho = \frac{\lambda_{tr} T}{4\pi\omega_p^2}
$$
would however predict a moderately larger value for hole- than for
electron-doped C$_{60}$, at least assuming (somewhat arbitrarily) the same
plasma frequencies for the same carrier densities.
Quantitatively, Batlogg's FET data (Sch\"on {\it et al.} 2000) differ
strongly from this expectation.
They, first of all, indicate a nonlinear temperature dependence, closer to
$T^2$; secondly, they show values about 5 times larger for holes than for
electrons.
While there are second order processes (see Appendix) that would
indeed yield a $T^2$ resistivity at low temperatures, we do not believe
that they may explain the discrepancy here.  
Zettl and coworkers (Vareka and Zettl 1994) proposed that the apparent
$T^2$ in the electron resistivity is an effect of thermal expansion, and
showed that a linear $T$ increase is recovered at constant volume, for
negative doping.

Recently Goldoni {\it et al.} measured by EELS the plasma frequency in
K$_3$C$_{60}$.
They found it slowly decreasing with temperature, its width growing
approximately quadratically with $T$.
These data support the view that the $T^2$ resistivity is directly related
with a $T^{-2}$ decrease of relaxation time, most likely linked with
lattice expansion.
It seems plausible that a similar physics could apply to holes too, in
which case the predicted constant-pressure relaxation-time drop with
temperature would also be non-linear, and quantitatively larger than the
electron case.
If, on the other hand, it became possible to obtain the constant-volume
inverse relaxation time and resistivity, then, assuming the same plasma
frequency $\omega_p$,  its increase should be
linear with $T$ with a slope 1.4 times larger than that of
negatively-charged C$_{60}$.
This conjecture must await experimental test.

\section*{Acknowledgements}

We are indebted to B.\ Batlogg, O.\ Gunnarsson and G.\ Santoro for useful
discussions.
This work was partly supported by the European Union, contract
ERBFMRXCT970155 (TMR Fulprop), and by MURST COFIN99.

\section*{Appendix: Quadratic Resistivity}

Within Fermi's golden rule, the vibron contribution to the electrical
resistivity is exponentially decreasing if temperature is much smaller than
the vibron frequencies. However, the above result is not true any more if
higher order corrections are taken into account.
Indeed, at second order, the electron-vibron coupling generates an
effective electron-electron interaction.
Since the electrons involved lie on a shell of width $T$ around the Fermi
energy, the vibron-originated electron-electron interaction
\begin{eqnarray}
V_{el-el}(\omega) &=& 
- \sum_{\tau,i,m} \frac{g_{\tau i}^2\hbar\omega_{\tau i}}{4}
\frac{\omega_{\tau i}^2}{\omega_{\tau i}^2 - \omega^2}\nonumber\\
&&\frac{1}{V}\sum_{kpq}\sum_{\mu\nu\gamma\beta}\sum_{\sigma,\sigma'}
V^{\tau(m)}_{\mu\nu}(\alpha) 
V^{\tau(m)}_{\gamma\beta}(\alpha) d^\dagger_{\mu,\sigma,k+q}
d^\dagger_{\gamma,\sigma',p}
d_{\beta,\sigma',p+q} d_{\nu,\sigma,k}\ ,
\label{M:Vel-el}
\end{eqnarray}
acts as if it were effectively unretarded, $\omega=0$. 
This interaction induces a $T^2$ inverse relaxation time at low temperatures. 
A rough estimation of the order of magnitude gives
\begin{eqnarray*}
\frac{1}{\tau} &\propto & \frac{(k_{\rm B} T)^2}{5\hbar} N_1(\epsilon_F)^3  
\sum_{i,j,\tau,\tau',m,m'}
\left(\frac{g_{\tau i}^2 \hbar\omega_{\tau i}}{4}\right) 
\left(\frac{g_{\tau' j}^2 \hbar\omega_{\tau' j}}{4}\right)\\ 
&& \left\{ \frac 54 
\left[{\rm Tr}\left(V^{\tau(m)}V^{\tau'(m')}\right)\right]^2 
- {\rm Tr}\left( V^{\tau(m)}V^{\tau'(m')}V^{\tau(m)}V^{\tau'(m')}\right)
\right\}\ .
\end{eqnarray*}

\bibliographystyle{prsty}

\end{document}